\documentclass[aps,pre,eqsecnum,showpacs,twocolumn]{revtex4}
\usepackage{amsmath,bm}

\begin{document}

\title{Passive Scalar Evolution in Peripheral Region}

\author{V. V. Lebedev$^{a,b}$ and K. S. Turitsyn$^a$}

\affiliation{$^a$ Landau Institute for Theoretical Physics,
Moscow, Kosygina 2, 117334, Russia \\
$^b$ Theoretical Division, LANL, Los Alamos, NM 87545, USA}
\date{\today}

\begin{abstract}

We consider evolution of a passive scalar (concentration of
pollutants or temperature) in a chaotic (turbulent) flow. A
universal asymptotic behavior of the passive scalar decay
(homogenization) related to peripheral regions (near walls) is
established. The passive scalar moments and its pair correlation
function in the peripheral region are analyzed. A special case
investigated in our paper is the passive scalar decay along a
pipe.


\end{abstract}

\maketitle

\section*{INTRODUCTION}

The problem of mixing attracts a great attention both due to its
fundamental significance and a variety of applications. Mixing
rate of an additive to a fluid is very sensitive to the character
of the hydrodynamic flow excited in the fluid. Recently an
essential breakthrough was achieved in the theory of the so-called
passive scalar (which can be concentration of pollutants or
temperature) in chaotic and turbulent flows (see reviews
\cite{00SS,01FGV}). The word ``passive'' means that a feedback of
the scalar field to the flow is negligible. It is correct for
dilute solutions of pollutants and for relatively weak
fluctuations of temperature. The passive scalar in a fluid is
subjected to diffusion (thermodiffusion) and advection, leading to
an evolution of its spatial distribution. A final result of the
evolution is a homogeneous state of the passive scalar. We are
interested in the passive scalar decay, that is in the laws
governing the homogenization of the passive scalar field in
chaotic and turbulent flows.

As it is noted in the paper \cite{03CLa} at large Peclet number Pe
the homogenization of the passive scalar field in the peripheral
regions is slower than in the bulk since mixing is suppressed near
walls of the vessel. Then the advanced stages of the passive
scalar decay are dominated just by the peripheral regions. In the
paper \cite{03CLa} main stages related to the passive scalar
evolution in the peripheral region were established (see also the
paper \cite{03CLb}). Here we develop further the theory of the
passive scalar evolution in the peripheral region having in mind
two different physical situations. The first case is the passive
scalar in the viscous boundary layer of the developed
(high-Reynolds) turbulence (explanations of its properties can be
found in the book of Monin and Yaglom \cite{MY}). The second case
is {a} chaotic flow. A perfect example of such flow is the
so-called elastic turbulence state revealed by Groisman and
Steinberg \cite{00GS} in dilute polymer solutions. Both the
velocity in the viscous boundary layer (in the first case) and the
chaotic velocity (in the second case) can be treated as smooth.
The smoothness of the random velocity field is one of key
ingredients of our analysis.

The theory of the passive scalar in a smooth random flow was
pioneered by Batchelor \cite{59Bat} and Kraichnan \cite{67Kra} who
considered the cases of velocities long-correlated and
short-correlated in time, respectively. The case of arbitrary
velocity correlation time was considered in the paper
\cite{95CFKLa}. In the works the passive scalar was assumed to
possess a stationary statistics due to pumping. Properties of the
passive scalar decay in an unbounded statistically homogeneous
turbulent media are also known. As it was demonstrated in the
paper \cite{00EX}, in the inertial range of the developed
turbulence the scalar decays in accordance with a power law.
Besides, it is interesting to consider the passive scalar decay in
the Batchelor region of scales (below the Kolmogorov viscous
scale), where the velocity field is smooth (for details see, e.g.,
the book of Batchelor \cite{Batchelor}). The decay law of the
passive scalar in the region is exponential, as it is demonstrated
in the papers \cite{99Son,99BF}. The same consideration is
applicable to the elastic turbulence. In the paper \cite{03CLa}
the combined case was considered, when both the inertial region of
scales and the Batchelor region of scales are taken into account.
Then the scalar decay is dominated by eddies from the inertial
interval (excluding, maybe, some initial stage of evolution) and
is, consequently, governed by a power law.

As it is noted in the paper \cite{03CLa} the theory of the passive
scalar in the random flow, developed for the bulk, needs a
modification for the peripheral region even for the smooth flow.
The reason is that the approximation of the velocity by linear
profiles, used for the bulk, fails in the peripheral region.
However, near a wall, the velocity possesses a definite dependence
on the separation from the wall, explaining the universal
properties of the passive scalar decay in the peripheral region.
Another property, simplifying an analysis, is the slowness of the
passive scalar mixing in the peripheral region, enabling one to
treat the velocity as short-correlated. The first problem, we
consider, is an evolution of the concentration of pollutants, when
the boundaries are considered as impenetrable for the pollutants.
This approach can be extended to the case of the binary chemical
reactions (see the paper \cite{03CLb}). Another problem, which can
be treated inside our approach, is the temperature relaxation in
{the} bulk, if the temperature at the walls of the vessel is
fixed. Then a heat flow from the boundary to the bulk is forced
{which} that is governed just by the velocity fluctuations in the
peripheral region. The passive scalar evolution can be examined
also in a fluid (where random motion is excited) flowing through a
pipe. Then the role of time is played by the coordinate along the
pipe. This setup is closely related to the experiment by Groisman
and Steinberg \cite{01GS}.

A remarkable property of a turbulent flow is its strong
intermittency leading to the so-called anomalous scaling of the
velocity correlation functions, which are dependent on the
integral scale of turbulence at a power, referred to as the
anomalous scaling index (see, e.g., the book of Frisch
\cite{Frisch}). Theoretically, an existence of the anomalous
scaling of such kind was established for the passive scalar in the
framework of the so-called Kraichnan model (see the papers
\cite{95KrM}). One could anticipate that the same phenomenon takes
place for the passive scalar decay. We demonstrate that in the
peripheral region the passive scalar moments possess, indeed, an
anomalous scaling, which is in some sense extreme. Namely, all the
moments of the passive scalar damp in accordance with the same
power law at increasing the separation from the wall.

The paper is organized as follows. In Section \ref{sec:general} we
present general equations, which describe the passive scalar
evolution in the peripheral region. In Section \ref{sec:periph} we
analyze the passive scalar evolution near the wall of the vessel.
In Section \ref{sec:pipe} we consider the passive scalar decay
along a pipe. Some general remarks and a short comparison with
experiment are presented in Conclusion.

\section{General Relations}
\label{sec:general}

Advection of a passive scalar field $\theta$ by a moving fluid
(accompanied by the passive scalar diffusion) is described by the
equation
 \begin{eqnarray}
 \partial_t\theta+\bm v\nabla\theta
 =\kappa\nabla^2\theta\,,
 \label{pe1} \end{eqnarray}
where $\bm v$ is the flow velocity and $\kappa$ is the diffusion
coefficient. Below, the fluid is assumed to be incompressible
(that is $\nabla\bm v=0$). A formal solution of the Cauchy problem
for the equation (\ref{pe1}) can be written as
 \begin{eqnarray}
 \theta(t_2)={\mathrm T}\exp\left\{
 \int_{t_1}^{t_2}\mathrm d t\,\left[
 -\bm v(t)\nabla +\kappa\nabla^2
 \right]\right\} \theta(t_1) \,,
 \label{pe2} \end{eqnarray}
where ${\mathrm T}\exp$ means a chronologically ordered exponent.
Of course, some boundary conditions for the passive scalar
$\theta$ should be introduced. There could be two different types
of the boundary conditions. If $\theta$ is temperature (and walls
are made of a well heat conducting material) then $\theta$ is
fixed at the boundary. If $\theta$ is density of pollutants and
the boundary is impenetrable for the pollutants then the gradient
of $\theta$ in the direction perpendicular to the boundary is zero
(that corresponds to zero pollutant flux to the boundary).

We consider a random flow which has to be characterized
statistically: via correlation functions. The correlation
functions are averages over time, they can be treated also as
averages over velocity realizations. The flow is assumed to be
statistically homogeneous in time, whereas there is no homogeneity
in space (because of the boundary effects). Here we consider the
flow in a closed vessel where the average velocity is equal to
zero. A generalization to the case when the average velocity is
non-zero is trivial (it can be found in Section \ref{sec:pipe}
where a fluid pushed through a pipe is treated). The pair velocity
correlation function $\langle v_\alpha(t_1,\bm r_1)
v_\beta(t_2,\bm r_2)\rangle$ depends on the time difference
$t_1-t_2$ only (due to the assumed time homogeneity), and on
coordinates of both points $\bm r_1$ and $\bm r_2$.

We have in mind two different situations. First, it is the viscous
boundary layer of the developed high-Reynolds turbulence (for
details see, e.g., the book \cite{MY}). Second, it is the
peripheral region of a chaotic flow. As a perfect example of the
chaotic flow, the elastic turbulence can be noted \cite{00GS}. In
both cases the velocity is smooth in the peripheral region we are
interested in. However, the velocity field possesses some
peculiarities, related to zero value of the velocity at the
boundary. That makes the passive scalar decay in the peripheral
region slow. Since the velocity correlation time is determined by
dynamics in the bulk, which is relatively fast, at examining the
passive scalar evolution in the peripheral region the velocity can
be treated as short correlated in time. It is well known, that in
this case closed equations for the passive scalar correlation
functions can be derived. Below, we demonstrate principal steps of
the derivation, based on Eq. (\ref{pe2}).

Let us examine the passive scalar evolution on a time interval
$(t_1,t_2)$ taking the difference $t_2-t_1$ much larger, than the
velocity correlation time $\tau$, but much smaller, than the
characteristic mixing time (the gap between the mixing time and
the velocity correlation time exists due to the noted weakness of
mixing in the peripheral region). The last condition enables one
to produce an expansion of the ${\mathrm T}$-exponent in Eq.
(\ref{pe2}). One may keep only two first terms of the expansion:
 \begin{eqnarray} &&
 \theta(t_2)\approx\theta(t_1)
 +(t_2-t_1)\kappa\nabla^2\theta(t_1)
 -\int_{t_1}^{t_2}\mathrm d t\,\bm v(t)\nabla\theta(t_1)
 \nonumber \\ &&
 +\int_{t_1}^{t_2}\mathrm dt\,
 \int_{t_1}^{t}\mathrm d t'
 \bm v(t)\nabla\left[\bm v(t')\nabla
 \theta(t_1)\right] \,.
 \label{pe3} \end{eqnarray}
The next step is in averaging over the velocity statistics inside
the interval $(t_1,t_2)$. This averaging is independent on the
velocity profiles at $t<t_1$ and $t>t_2$ due to the condition
$t_2-t_1\gg\tau$. Averaging the expression (\ref{pe3}), one
obtains a relation for the average value $\langle\theta\rangle$ of
the passive scalar
 \begin{eqnarray}
 \langle\theta(t_2,\bm r)\rangle-
 \langle\theta(t_1,\bm r)\rangle
 =(t_2-t_1)\kappa\nabla^2
 \langle\theta(t_1,\bm r)\rangle
 \nonumber \\
 +(t_2-t_1)\nabla_\alpha
 \left[D_{\alpha\beta}(\bm r,\bm r)
 \nabla_\beta\langle\theta(t_1,\bm r)\rangle \right] \,,
 \label{pe4} \\
 D_{\alpha\beta}(\bm r_1,\bm r_2)=
 \int_0^\infty\mathrm d t\,
 \langle v_\alpha(t,\bm r_1)
 v_\beta(0,\bm r_2)\rangle \,
 \label{pe5} \end{eqnarray}
where a fast enough decay of the pair velocity correlation
function with $t$ is implied. We used also the incompressibility
condition $\nabla\bm v=0$. The quantity $\langle\theta\rangle$
means a value, averaged over the velocity fluctuations. The
quantity $D_{\alpha\beta}(\bm r,\bm r)$, entering Eq. (\ref{pe4}),
is usually called the turbulent diffusion tensor. If $t_2-t_1$ is
smaller than the mixing time, then the right-hand side of the
equation (\ref{pe4}) is a small correction to
$\langle\theta\rangle$. Therefore the equation (\ref{pe4}) can be
rewritten in the differential form:
 \begin{eqnarray}
 \partial_t\langle\theta\rangle=\nabla_\alpha
 \left[D_{\alpha\beta}(\bm r,\bm r)
 \nabla_\beta\langle\theta\rangle\right]
 +\kappa\nabla^2\langle\theta\rangle \,.
 \label{pe6} \end{eqnarray}

Analogously, starting from Eq. (\ref{pe2}), one can derive closed
equations for high-order correlation functions of $\theta$. Say,
the equation for the pair correlation function $F$ is
 \begin{eqnarray} &&
 \partial_tF(t,\bm r_1,\bm r_2)
 =\kappa(\nabla_1^2+\nabla_2^2)F
 \label{pe8} \\ &&
 +\nabla_{1\alpha}
 \left[D_{\alpha\beta}(\bm r_1,\bm r_1)
 \nabla_{1\beta}F\right]
 +\nabla_{2\alpha}
 \left[D_{\alpha\beta}(\bm r_2,\bm r_2)
 \nabla_{2\beta}F\right]
 \nonumber \\ &&
 +\nabla_{1\alpha}
 \left[D_{\alpha\beta}(\bm r_1,\bm r_2)
 \nabla_{2\beta}F\right]
 +\nabla_{2\alpha}
 \left[D_{\alpha\beta}(\bm r_2,\bm r_1)
 \nabla_{1\beta}F\right] \,,
 \nonumber \\ &&
 F(t,\bm r_1,\bm r_2)=
 \langle\theta(t,\bm r_1)
 \theta(t,\bm r_2)\rangle \,.
 \label{pe7} \end{eqnarray}
Generally, the equation for the $n$-th order correlation function
$F_n$ of the passive scalar is
\begin{eqnarray} &&
 \partial_tF_n
 =\kappa\sum_{m=1}^n\nabla_m^2F_n
 \nonumber \\ &&
 +\sum_{m,k=1}^n\nabla_{m\alpha}
 \left[D_{\alpha\beta}(\bm r_m,\bm r_k)
 \nabla_{k\beta}F_n\right]\,,
 \label{pe10} \end{eqnarray}
 \begin{equation}
  F_n(t,\bm r_1,\dots,\bm r_n)=
 \langle\theta(t,\bm r_1)\dots
 \theta(t,\bm r_n)\rangle \,.
 \label{pe9} \end{equation}
The structure of the equation (\ref{pe10}) is transparent: the
evolution of the passive scalar correlation function is determined
by the molecular diffusion (the first term in the right-hand side)
and by the turbulent diffusion (the second term in the right-hand
side).

Note that if the diffusivity is negligible then it is possible to
obtain a closed equation for the moments $\langle\theta^n(\bm
r)\rangle=F_n(\bm r,\dots,\bm r)$ from Eq. (\ref{pe10})
 \begin{eqnarray}
 \partial_t\langle\theta^n(\bm r)\rangle
 =\nabla_\alpha\left[D_{\alpha\beta}(\bm r,\bm r)
 \nabla_\beta\langle\theta^n(\bm r)\rangle\right]  \,,
 \label{pe66} \end{eqnarray}
which is identical to the equation (\ref{pe6}), without the molecular
diffusion term. Eq. (\ref{pe66}) is a direct consequence of the
relation $\partial_t(\theta^n)=-\bm v\nabla(\theta^n)$, which
follows from Eq. (\ref{pe1}), if the molecular diffusion is neglected.

The turbulent diffusion tensor $D$ can be estimated as $D\sim
V^2\tau$, where $V$ is characteristic value of the velocity
fluctuations, and $\tau$ is the velocity correlation time, that
can be estimated as $\tau\sim L/V_L$. Here $L$ is the size of the
viscous boundary layer in the case of the high-Reynolds turbulence
and the size of the vessel in the case of the chaotic flow
(elastic turbulence), and $V_L$ is characteristic value of the
velocity fluctuations in the bulk. Therefore, the weakness of the
passive scalar decay is explained by smallness of the ratio
$V/V_L$ inside the peripheral region. Note that for the elastic
turbulence $\tau$ is determined by the polymer relaxation time and
the condition $\tau\sim L/V_L$ is no other than the Lumley
criterion of strong polymer elongation formulated in the papers
\cite{73Lum} (see also the papers \cite{BFL}).

Let us explain physical meaning of the passive scalar correlation
functions. They are averages over the velocity statistics.
Therefore, to obtain the correlation functions experimentally or
in numerics, one has to measure the passive scalar decay many
times (for many realizations of the velocity field) and then to
average the result over the attempts. Initial conditions for the
passive scalar field are implied to be fixed at the averaging
procedure. If we consider the case of a fluid pushed through a
pipe, then the passive scalar correlation functions are
stationary. Then they can be treated as averages over long time.

Below we consider a passive scalar evolution in the peripheral
region. We assume that mixing already produced the homogeneous
distribution of the passive scalar in the bulk (remind that the
process in the bulk is much faster than in periphery). And we
subtract from $\theta$ a constant, corresponding to the bulk value
of $\theta$ (this redefinition does not change the equations
describing the passive scalar evolution). By other words, the
value of $\theta$ tends to zero when we go away from the boundary.
Next, the homogenization of the passive scalar along the boundary
is much more effective than its homogenization in the direction
perpendicular to the wall. Therefore, analyzing the evolution of
the passive scalar after the homogenization in the bulk, we can
treat its initial condition $\theta_0$ as dependent mainly on the
separation from the wall and practically independent of
coordinates along the wall.

\section{Decay in Vessel}
\label{sec:periph}

Here we consider the passive scalar decay in a closed vessel. If
the walls of the vessel are smooth and their curvature is of the
order of the vessel size, then at considering the peripheral
region the boundary can be treated as flat in the main
approximation. Then it is possible to introduce the orthogonal
reference system, where the coordinate $q$ measures separation
from the boundary and $r_\mu$ are coordinates along the wall, the
subscript $\mu$ ($\nu$) running over 1 and 2. In the reference
frame the incompressibility condition is written as $\partial_q
v_q+\partial v_1/\partial r_1+\partial v_2/\partial r_2=0$, where
$v_q$ is the velocity component perpendicular to the wall and
$v_\mu$ are {the} components along the wall. Since $v_\mu$ tends
to zero as $q\to0$ then $v_\mu\propto q$ near the boundary, and
incompressibility leads to the proportionality law $v_q\propto
q^2$. This is the main feature of the velocity in the peripheral
region.

For the flat wall, it is naturally to assume homogeneity of the
velocity correlation functions along the wall and also their
isotropy in the planes parallel to the wall. Then we obtain a
general expression for the components of the turbulent diffusivity
tensor (\ref{pe5})
 \begin{eqnarray} &&
 D_{\mu\nu}(\bm r_1,\bm r_2)
 =H_1(\varrho)\delta_{\mu\nu} q_1 q_2
 +H_2(\varrho) \varrho_\mu \varrho_\nu q_1 q_2 ,
 \nonumber \\ &&
 D_{q\nu}(\bm r_1,\bm r_2)
 =-\frac{1}{2}H_1'(\varrho)\frac{\varrho_\nu}{\varrho}q_1^2q_2
 \nonumber \\ &&
 -\frac{1}{2}\left[\varrho H_2'(\varrho)
 +3 H_2(\varrho)\right]{\varrho_\nu}q_1^2q_2 ,
 \nonumber \end{eqnarray}
 \begin{eqnarray} &&
 D_{\mu q}(\bm r_1,\bm r_2)
 =\frac{1}{2}H_1'(\varrho)\frac{\varrho_\mu}{\varrho}q_1q_2^2
 \nonumber \\ &&
 +\frac{1}{2}\left[\varrho H_2'(\varrho)
 +3 H_2(\varrho)\right]{\varrho_\mu}q_1q_2^2 ,
 \nonumber \\ &&
 D_{qq}(\bm r_1,\bm r_2)=
 -\frac{1}{4}\left[\frac{H_1'(\varrho)}{\varrho}
 +H_1''(\varrho)\right]q_1^2 q_2^2
 \nonumber \\ &&
 -\frac{1}{4}\left[6H_2(\varrho)+6\varrho H_2'(\varrho)
 +\varrho^2 H_2''(\varrho)\right]q_1^2 q_2^2 ,
 \label{pr1} \end{eqnarray}
where $\varrho_\mu=r_{1\mu}-r_{2\mu}$ and $H_1$, $H_2$ are some
functions of $\varrho$. The expressions (\ref{pr1}) satisfy the
conditions $\nabla_{1\alpha}D_{\alpha\beta}(\bm r_1,\bm r_2)=0=
\nabla_{2\beta}D_{\alpha\beta}(\bm r_1,\bm r_2)$, following from
the incompressibility condition.

Since we consider the peripheral region where the velocity field
is smooth, then both $H_1$ and $H_2$ have regular expansion in
$\varrho$ (containing even powers) at small $\varrho$, that is at
$\varrho\ll L$ (remind that $L$ is the thickness of the peripheral
region which is the thickness of the viscous boundary layer in the
case of the developed turbulence and is of the order of the vessel
size for the elastic turbulence). In this limit the main
contribution to the turbulent diffusion tensor is determined by
first terms of the expansion:
 \begin{eqnarray} &&
 H_1\approx H_{10}-(\mu+3H_{20}/2)\varrho^2, \qquad
 H_2\approx H_{20},
 \label{pr2} \\ &&
 D_{\mu\nu}\approx H_{10} \delta_{\mu\nu}q_1q_2, \qquad
 D_{qq}\approx \mu q_1^2 q_2^2 .
 \label{pr3} \end{eqnarray}
The quantity $\mu$ characterizes the flow intensity near the
boundary. It can be estimated as $\mu\sim V_L/L^3$, where, as
previously, $V_L$ is the characteristic velocity fluctuation in
the bulk. In the framework of the Karman-Prandtl theory of the
viscous boundary layer (for details see, e.g., the book \cite{MY})
the width of the layer is estimated as $L\sim \nu/V_L$ (where
$\nu$ is kinematic viscosity of the fluid) and one finds $\mu\sim
V_L^4\nu^{-3}$. For the elastic turbulence $L$ is the vessel size
and $\mu\sim V_L^4\nu^{-3}\mathrm{Re}^{-3}$, where $\mathrm{Re}=
V_LL/\nu$ is the Reynolds number.

Comparing the turbulent diffusion term for the motion
perpendicular to the wall, which can be estimated as $\mu
q^4\partial_q^2$, and the molecular diffusion term
$\kappa\partial_q^2$, one finds the width of the boundary
diffusion layer
 \begin{equation}
 r_{bl}=(\kappa/\mu)^{1/4}.
 \label{rbl} \end{equation}
We assume that $r_{bl}\ll L$. The relation can be rewritten as
$\mathrm{Pe}\gg1$ where $\mathrm{Pe}$ is the Peclet number:
$\mathrm{Pe}=V_L L/\kappa$. As it follows from Eq. (\ref{rbl}),
$L/r_{bl}\sim \mathrm{Pe}^{1/4}$. For the case of the viscous
boundary layer, one obtains $\mathrm{Pe}\sim\mathrm{Sc}$, where
$\mathrm{Sc}=\nu/\kappa$ is the Schmidt number. For the elastic
turbulence, one finds a slightly different estimate $\mathrm{Pe}
\sim\mathrm{Sc}\cdot\mathrm{Re}$.

 \subsection{Average Scalar}
 \label{subsec:average}

Here we consider the simplest possible correlation function: the
average value of the passive scalar $\langle\theta\rangle$. The
quantity essentially depends on the separation from the wall $q$
and slowly depends on the coordinates along the wall. We ignore
the last dependence, assuming that its characteristic length is
the vessel size. Then we find from Eqs. (\ref{pe6},\ref{pr3}) the
following equation for the object
 \begin{eqnarray}
 \partial_t\langle\theta\rangle
 =\mu\partial_q\left(q^4
 \partial_q\langle\theta\rangle\right)
 +\kappa\partial_q^2\langle\theta\rangle \,.
 \label{pe11} \end{eqnarray}
The advection term (with $\mu$) in Eq. (\ref{pe11}) dominates at
$q\gg r_{bl}$ and the diffusion term (with $\kappa$) dominates at
$q\ll r_{bl}$, where $r_{bl}$ is thickness of the boundary
diffusion layer defined by Eq. (\ref{rbl}). As we explained in the
last paragraph of Section \ref{sec:general}, at solving the
problem we should assume $\lim_{q\to\infty}\langle\theta\rangle
=0$, since large $q$ correspond to the bulk, where $\theta$ is
assumed to tend to zero. The condition is implied below.

We examine the passive scalar evolution in the peripheral region,
which begins after that its homogenization in the bulk is
finished. Then the initial distribution of the passive scalar has
the characteristic length $L$. The subsequent evolution is divided
into two stages. At the first stage the thickness $\delta$ of the
layer, where $\theta$ is concentrated, diminishes as $\delta=(\mu
t)^{-1/2}$. When $\delta$ reaches $r_{bl}$, the second stage
starts, which is characterized by the fixed spatial scale
$r_{bl}$.

At treating the first stage one can omit the diffusion term (with
$\kappa$) in Eq. (\ref{pe11}), that leads to the equation
\begin{eqnarray}
 \partial_t\langle\theta\rangle
 =\mu\partial_q\left(q^4
 \partial_q\langle\theta\rangle\right) \,.
 \label{rbl1} \end{eqnarray}
Looking for a solution $\langle\theta\rangle=
\exp(-st)\varphi_s(q)$, one obtains from Eq. (\ref{rbl1})
 \begin{eqnarray}
 \varphi_s =\sqrt{\frac{\pi}{2}}\,
 \left(\frac{s}{\mu q^2}\right)^{3/4}
 J_{3/2}\left(\sqrt{s/\mu}\, q^{-1} \right) \,.
 \label{pe12} \end{eqnarray}
Using the orthogonality relation
 \begin{eqnarray}
 \int_0^\infty\mathrm dq\,
 \varphi_s(q)\varphi_\sigma(q)
 =\frac{\pi s^{3/2}}{\mu^{1/2}}
 \delta(s-\sigma) \,,
 \nonumber \end{eqnarray}
one finds a general solution of Eq. (\ref{rbl1})
 \begin{equation}
 \langle\theta(t,q)\rangle
 =\int_0^\infty\!\!\!\!\mathrm d s\,\varphi_s(q)
 \frac{\sqrt\mu\,e^{-st}}{\pi s^{3/2}}
 \int_0^\infty\!\!\!\! \mathrm dq'\,\varphi_s(q')
 \theta_0(q')
 \label{pp14} \end{equation}
in terms of the initial passive scalar distribution. The
approximation (\ref{pp14}) is correct provided $\delta\gg r_{bl}$
(then it is possible to neglect the diffusion boundary layer where
diffusion is relevant).

We start from a distribution of $\theta$ with the characteristic
length $L$. Therefore at times, when $\delta\ll L$  and at $q\ll
L$ one can substitute $\theta(t=0)$ in Eq. (\ref{pp14}) by
$\vartheta_0=\theta(t=0,q=0)$. Taking then the integrals in Eq.
(\ref{pp14}), one derives
 \begin{eqnarray} &&
 \langle\theta(t,q)\rangle=
 \frac{\vartheta_0}{\pi}
 \int_0^\infty\frac{\mathrm d s}{s}
 e^{-st}\varphi_s(q) \,.
 \nonumber \\ &&
 =\vartheta_0\left[\mathrm{erf}\left(\frac{\delta}{2q}\right)
 -\frac{\delta}{\sqrt\pi\,q}
 \exp\left(-\frac{\delta^2}{4 q^2}\right)\right] \,.
 \label{pe20} \end{eqnarray}
This profile has a universal form insensitive to
details of the initial distribution of $\theta$.
If $q\gg\delta$ then one obtains from Eq. (\ref{pe20})
 \begin{eqnarray}
 \langle\theta\rangle\approx
 \frac{\vartheta_0 \delta^3}{6\sqrt\pi q^3} \,.
 \label{pp15} \end{eqnarray}
If $q\ll\delta$ then we
find $\langle\theta\rangle=\vartheta_0$. So, the value of
$\theta$ is practically unchanged inside the layer $q<\delta$.

Note that though the equation (\ref{rbl1}) has form of the
conservation law for $\langle\theta\rangle$, the total amount of
the passive scalar in the peripheral region $\int\mathrm
dq\,\langle\theta\rangle$ appears to be time dependent, if the
expression (\ref{pe20}) is integrated. The reason is that the
considered solution corresponds to non-zero passive scalar flux
directed to large $q$, i.e. to the bulk, which can be treated as a
big reservoir. This flux $\mu q^4 \partial_q \langle\theta\rangle$
can be obtained directly from the asymptotic expression
(\ref{pp15}). Note also that the passive scalar evolution at the
first stage is insensitive to the boundary conditions, and
therefore it is described identically for the concentration of
pollutants and temperature.

Now we analyze the passive scalar behavior at the second stage,
then the diffusion term can not be ignored. Let us first examine
the case when the passive scalar represents the concentration of
pollutants. Then the equation (\ref{pe11}) has to be supplemented
by the boundary condition $\partial_q \langle\theta\rangle =0$ at
$q=0$, which means zero flux of pollutants to the boundary. At
long times, only the contribution related to the minimal (by its
absolute value) eigen value of the operator in the right-hand side
of Eq. (\ref{pe11}) is left. That leads to the exponential decay
$\langle\theta\rangle\propto \exp(-\gamma t)$, the decrement
$\gamma$ is related to the energy of the ground state. The value
of $\gamma$ can be found numerically, it is $\gamma=
c_E\sqrt{\kappa\mu}$, $c_E \approx 1.81$. The asymptotic behavior
of $\langle\theta\rangle$ can be related to the initial value of
the passive scalar $\vartheta_0$ near the wall:
$\langle\theta(q=0)\rangle=c_0\vartheta_0\exp(-\gamma t)$, where
$c_0 \approx 1.55$. The total amount of the scalar near the
boundary behaves as $\int\mathrm{d}q\,\langle \theta\rangle =c_1
\vartheta_0 r_{bl} \exp(-\gamma t)$, where $c_1 \approx 1.55$.

Let us now consider the case when the passive scalar $\theta$
represents temperature, assuming that it is fixed at the boundary:
$\theta(q=0)=\vartheta_0$. Then after the first stage a
quasi-stationary distribution of $\langle\theta\rangle$ is formed,
since the bulk can be treated as a big reservoir having a constant
temperature. This quasi-stationary distribution can be found
directly from the equation (\ref{pe11}) where the term with the
time derivative has to be omitted:
 \begin{equation}
 \langle\theta\rangle=
 \frac{2\sqrt2}{\pi}\,\kappa^{3/4}\mu^{1/4}\vartheta_0
 \int_q^\infty\frac{\mathrm d q_1}{\mu q_1^4+\kappa} .
 \label{rbl3} \end{equation}
At $q\gg r_{bl}$ we find, again, $\langle\theta\rangle\propto
q^{-3}$. That corresponds to a non-zero passive scalar flux (heat
flux) to the bulk. This flux is time-independent in the case
{considered}.

 \subsection{High moments}
 \label{subsec:scalarpdf}

As we already noted, at the first stage the diffusive term in the
equation for the passive scalar correlation functions can be
omitted. Then the closed equation (\ref{pe66}) for high moments of
the passive scalar is correct. The high moments (similar to the
first one) depend mainly on $q$ and we ignore their dependence on
the coordinates along the wall. Then, substituting into the
equation the expressions (\ref{pr3}), one finds
 \begin{eqnarray}
 \partial_t\langle\theta^n\rangle
 =\mu\partial_q\left(q^4
 \partial_q\langle\theta^n\rangle\right) \,,
 \label{pr4} \end{eqnarray}
which is a generalization of Eq. (\ref{rbl1}). Its solution can be
written as Eq. (\ref{pp14})
 \begin{equation}
 \langle\theta^n\rangle
 =\int_0^\infty\!\!\!\!\mathrm d s\,\varphi_s(q)
 \frac{\sqrt\mu\,e^{-st}}{\pi s^{3/2}}
 \int_0^\infty\!\!\!\! \mathrm dq'\,\varphi_s(q')
 \theta_0^n(q') .
 \label{pr14} \end{equation}
The expression (\ref{pr14}) implies that $\delta\gg r_{bl}$ since
only at this condition it is possible to neglect the region $q\sim
r_{bl}$ where the diffusion is relevant. Since the initial
distribution of $\theta$ has the characteristic length $L$, at the
condition $\delta\ll L$ we obtain, again, a universal expression
 \begin{equation}
 \langle\theta^n\rangle
 =\vartheta_0^n\left[\mathrm{erf}\left(\frac{\delta}{2q}\right)
 -\frac{\delta}{\sqrt\pi\,q}
 \exp\left(-\frac{\delta^2}{4q^2}\right)\right],
 \label{pr20} \end{equation}
which is a generalization of Eq. (\ref{pe20}). The expression
shows that in the region $q\gg\delta$, $\langle\theta^n\rangle
\approx\vartheta_0^n\delta^3/(6\sqrt\pi\,q^3)$.

Actually, the expression (\ref{pr14}) is correct for any averaged local
function of the scalar $\theta$. We can use it to find the local scalar PDF
$P(t,q,\theta)=\langle\delta\left[\theta-\theta(t,q)\right]\rangle$.
Let us first rewrite the expression like (\ref{pr14}) as
 \begin{eqnarray} &&
 P(t,q)= \frac{1}{q^{3/2}}\int_0^\infty\mathrm dk\,
 \int_0^\infty\frac{\mathrm d q'}{(q')^{3/2}}\,
 k\exp(-\mu k^2t)
 \nonumber \\ && \times
 J_{3/2}(k/q) J_{3/2}(k/q')
 P(0,q') \,,
 \label{pe24} \end{eqnarray}
One can take the integral over $k$ in Eq. (\ref{pe24}) to obtain
 \begin{eqnarray} &&
 P = \frac{1}{\sqrt{\pi \mu t}}
 \int_0^\infty \frac{\mathrm{d}q'}{qq'}
 \exp\left[-\frac{1}{4 q^2 \mu t}-\frac{1}{4 q'^2 \mu t}\right] P(0,q')
 \nonumber \\  && \times
 \left\{\cosh\left[\frac{1}{2q q'\mu t}\right]
 - 2\mu t q q'
 \sinh\left[\frac{1}{2 q q'\mu t}\right]\right\} \,.
 \label{pdf2} \end{eqnarray}
Let us assume that the initial scalar distribution is a monotonic
function $\theta_0(q)$, which is equal to zero at $q\to\infty$ and
reaches a maximum value at $q\to0$. Then $P(t=0,q,\theta) =
-[1/\theta_0'(q)] \delta[q-q_0(\theta)]$, where the function
$q_0(\theta)$ is determined from the relation $\theta_0(q_0)
=\theta$. Calculating the integral (\ref{pdf2}) one finds
 \begin{eqnarray} &&
 P(t,q,\theta) = \frac{1}{q q_0|\theta_0'(q_0)|}
 \biggl[g\left(\frac{1}{q}-\frac{1}{q_0}\right)
 \left(1- 2\mu t q q_0\right)
 \nonumber \\ &&
 +g\left(\frac{1}{q}+\frac{1}{q_0}\right)\left(1+ 2\mu t q
 q_0\right)\biggr]\,,
 \label{pdf3} \\ &&
 g(x) = \frac{1}{2\sqrt{\pi \mu t}}
 \exp\left(-\frac{x^2}{4\mu t}\right) \,.
 \nonumber \end{eqnarray}
When $t$ grows, the boundary layer, determined by $\delta$,
shrinks, that is the characteristic value of $q$ decreases.
Besides, $q_0$ is fixed at a given $\theta$. Therefore one can
expand the expression in the square brackets in Eq. (\ref{pdf3})
in $1/q_0$ to obtain a universal probability distribution
 \begin{eqnarray}
 P=\frac{1}{12 q^2 \sqrt\pi\,
 (\mu t)^{5/2}q_0^3|\theta'_0(q_0)|}
 \exp\left(-\frac{1}{4 q^2 \mu t}\right) \,.
 \label{pdf33} \end{eqnarray}
Unfortunately, the expression (\ref{pdf33}) cannot be utilized for
calculation of the moments of $\theta$, since the integrals
$\int\mathrm d\theta\, \theta^n P(\theta)$ diverge near the
maximum value of $\theta$, $\vartheta_0$. The reason is that the
expression (\ref{pdf33}) is correct only if $q_0(\theta)\gg q$,
which is violated at small $q_0$, corresponding to $\theta$ close
to $\vartheta_0$.

At the second stage diffusion starts to be relevant, and it is
impossible to obtain closed equations for the moments
$\langle\theta^n(t,q)\rangle$. To find the moments one has to
solve the complete equations (\ref{pe10}) for the passive scalar
correlation functions, that is a complicated problem. One can say
only that due to linearity of the problem, the asymptotic in time
behavior of the correlation functions (and, consequently, moments)
is determined by the minimal eigen value of the operator in the
right-hand side of the equation (\ref{pe10}). Therefore for the
concentration of pollutants $\langle\theta^n(t,q)\rangle
\propto\exp(-\gamma_n t)$ where $\gamma_n\sim \sqrt{\kappa\mu}$.
The estimate can be obtained by equating the time derivative, the
turbulent diffusion term and the molecular diffusion term in Eq.
(\ref{pe10}). If we consider the region $q\gg r_{bl}$ then it is
possible to neglect the molecular diffusion term in Eq.
(\ref{pe10}) and we return to the closed equation (\ref{pr4}).
There the time derivative $\partial_t$ can be substituted by
$-\gamma_n$ and we conclude that at the same condition $q\gg
r_{bl}$ it is possible to neglect the term with the time
derivative in Eq. (\ref{pr4}). Therefore we obtain, again, the
asymptotic behavior
 \begin{eqnarray} &&
 \langle\theta^n(t,q)\rangle\propto q^{-3}\,,
 \label{anom} \end{eqnarray}
at $q\gg r_{bl}$. The situation here is analogous to one for the
average passive scalar, since the equation (\ref{pr4}) has the
form of conservation law for the high moments. The laws
$\langle\theta^n(t,q)\rangle\propto q^{-3}$ correspond to non-zero
fluxes of high degrees of the passive scalar.

Now we can turn to the case when the passive scalar is
temperature, fixed at the boundary. If the temperature in the bulk
is different from that at the boundary then a heat flow is
produced from the boundary to the bulk. Since the bulk is a big
reservoir then in the main approximation its temperature can be
treated as time-independent. In this case statistics of $\theta$
becomes quasistationary at the second stage (all the correlation
functions are independent of time). The equations for the passive
scalar correlation functions are determined by Eq. (\ref{pe10})
where the time derivative can be omitted. Particularly, at $q\gg
r_{bl}$ one obtains the closed equation for the moments
(\ref{pr4}) (where time derivative has to be omitted) leading to
the same law (\ref{anom}). Inside the diffusion boundary layer
$\theta\sim\vartheta_0$, where $\vartheta_0$ is the temperature at
the boundary (remind that we assume $\theta=0$ in the bulk). Then,
as it follows from the equation (\ref{pe66}),
$\langle\theta^n\rangle\sim\vartheta_0^n (r_{bl}/q)^3$.

We established that if $q\gg\delta$ at the first stage or if $q\gg
r_{bl}$ at the second stage then the law (\ref{anom}) is valid.
Moreover, it is possible to formulate the estimates
$\langle\theta^n\rangle\sim \langle\theta\rangle^n
(q/\delta)^{3(n-1)}$ for the first stage and the inequalities
$\langle\theta^n\rangle\gtrsim \langle\theta\rangle^n
(q/r_{bl})^{3(n-1)}$ for the second stage. The estimates show that
the high moments of the passive scalar are much larger than their
Gaussian evaluation, this property implies strong intermittency in
the system. The expression (\ref{anom}) can be treated as a
manifestation of anomalous scaling. It is extreme since the
exponent characterizing $q$-dependence of the moments is
independent of $n$.

\subsection{Pair Correlation Function}

Here we examine the pair correlation function $F(\bm r_1,\bm
r_2)$. One could anticipate its non-trivial scaling behavior in
the region $q_1,q_2\gg \delta$ (for the first stage) or
$q_1,q_2\gg r_{bl}$ (for the second stage). If $q_1\gg q_2$ then
the only term $\nabla_{1\alpha}[D_{\alpha\beta}(\bm r_1,\bm
r_1)\nabla_{1\beta}F]$ in Eq. (\ref{pe8}) should be kept, and we
come to the conclusion that $F\propto q_1^{-3}$. Therefore we
obtain the behavior, which is similar to the passive scalar
moments in this asymptotic case.

Now we consider the case of close points $\bm r_1$ and $\bm r_2$,
assuming small value of $\varrho_\mu=r_{1\mu}-r_{2\mu}$,
$\varrho\ll L$, admitting the expressions (\ref{pr3})
and $q\ll Q$, where $Q=(q_1+q_2)/2$ and $q=q_1-q_2$.
Then one obtains from Eq. (\ref{pe8})
 \begin{eqnarray} &&
 \partial_tF=\mu\partial_Q (Q^4 \partial_Q F)
 + (\kappa/2)\partial_Q^2 F +2 \kappa \partial_q^2 F
 \nonumber \\ &&
 +\Bigl\{4\mu Q^3q\partial_Q\partial_q
 -2\mu Q^3 \varrho \partial_Q\partial_\varrho
 \nonumber \\ &&
 +\left[H_{10}q^2/\varrho-2(2\mu+3 H_{20})Q^2
 \varrho\right]\partial_\varrho \Bigr\} F
 \nonumber \\ &&
 +\Bigl\{4\mu Q^2q^2\partial_q^2-2H_{30}Q^4\varrho^2\partial_q^2
 -2\mu Q^2 \varrho q \partial_q\partial_\varrho
 \nonumber \\ &&
 +(2\mu+H_{20})Q^2\varrho^2 \partial_\varrho^2
 +H_{10}q^2\partial_\varrho^2\Bigr\} F .
 \label{pr6} \end{eqnarray}
Details of the derivation of Eq. (\ref{pr6}) can be found in
Appendix \ref{pairderiv}. It is natural to write $F=
\left\langle\theta^2(Q)\right\rangle\,(1+\varsigma)$ Here unity
represents the main contribution to the pair correlation function,
related to the second-order moment, and $\varsigma$ is a small
correction tending to zero at $\bm r_1\to\bm r_2$.

In the region $Q\gg\delta$ for the first stage or in the region
$Q\gg r_{bl}$ for the second stage the time derivative in Eq.
(\ref{pr6}) can be omitted in comparison with the turbulent
diffusion term, which is the first one in the right-hand side of
Eq. (\ref{pr6}). If the molecular diffusion terms in Eq.
(\ref{pr6}) are also neglected then one finds
 \begin{eqnarray} &&
 \Bigl\{(2 H_{30}\varrho^2-7 x^2)\partial_x^2-11 x\partial_x +3 +
 \nonumber \\ &&
 \left[(2+H_{20})\varrho^2+H_{10}x^2\right]\partial_\varrho^2+
 \nonumber \\ &&
 \left(2(1-3 H_{20})\varrho+H_{10}x^2/\varrho\right)
 \partial_\varrho\Bigr\}\varsigma = 0,
 \label{ss4} \end{eqnarray}
where $x=q/Q$ and we have taken into account that
$\left\langle\theta^2(Q)\right\rangle\propto Q^{-3}$. A solution
of the equation (\ref{ss4}) can be written as a sum of
$\varsigma_b=(qQ)^b f_b(x,\varrho)$ where $f_b$ satisfy the
equation (\ref{ss4}) (since $qQ$ is zero mode of the operator
figuring in this equation). It is clear that the main contribution
is related to $b=0$ (since negative $b$ are forbidden). Since the
operator in Eq. (\ref{ss4}) has definite scaling properties,
solutions of the equation can be written in a simple self-similar
form $\varsigma=(q/Q)^a\Psi(Q\varrho/q)$. The principal
contribution to $\varsigma$ is associated with the smallest $a$.
Unfortunately, the exponent $a$ is non-universal being dependent
on the coefficients $H_{ij}$. The molecular diffusion smoothes the
function $\varsigma$ at $q\sim r_{bl}^2/Q$.

The above expressions demonstrate the following natural behavior
of the pair correlation function. The main anomalous dependence of
the function is related to the second moment of the passive
scalar. As to the dependence on the relative separation between
the points $\bm r_1$ and $\bm r_2$, it is described in terms of
the function $\varsigma$ possessing a non-trivial scaling behavior
with a self-similar factor depending on the combination
$Q\varrho/q$.

\section{Decay along pipe}
\label{sec:pipe}

Here we discuss the case when the passive scalar decays along a
pipe in a statistically homogeneous flow, which is assumed to be
chaotic and has an average velocity $u$ along the pipe. Such setup
was used by Groisman and Steinberg in their experiments
\cite{01GS}. In this case the chaotic flow (elastic turbulence)
was excited in a polymer solution pushed through a curvilinear
pipe. The scalar dynamics is then governed by the equation
 \begin{equation} \label{scalarpipe}
 \partial_t \theta + u \partial_z \theta + v_\alpha
 \nabla_\alpha \theta = \kappa \nabla^2 \theta,
 \end{equation}
where $\bm v$ is fluctuating part of the velocity (with zero mean)
and $z$ is coordinate along the pipe. Below, the average velocity
$u$ is assumed to be much larger, than $\bm v$ (that corresponds
to the experimental situation).

If the pressure difference, pushing the flow, is constant, the
flow is statistically stationary and homogeneous along the pipe
(which is assumed to have a constant cross-section). Therefore $u$
is independent of $z$ and the velocity correlation functions do
depend on the time differences and coordinate differences along
the pipe. Thus, for a problem with a stationary scalar injection
to the pipe the scalar statistics is time-independent and the
coordinate $z$ along the pipe plays the role of time in the decay
problem. Following the procedure described in the first section
one obtains from Eq. (\ref{scalarpipe}) the following equations
for the scalar correlation functions
 \begin{equation}
 u \partial_z F_n = \kappa \sum_{m=1}^n \nabla^2_m F_n +
 \sum_{m,k=1}^n
 \nabla_{m\alpha}\left[D_{\alpha\beta}({\bm r}_m,{\bm r}_k)
 \nabla_{k\beta}\right] F_n.
 \label{pipen} \end{equation}
The only difference in comparison with Eq. (\ref{pe10}) is that
the time derivative is substituted by the advection term along the
pipe.

As previously, we introduce the coordinate $q$ measuring a
separation of a given point from the wall. The average velocity
$u$ is a function of $q$, tending to zero as $u \propto q$ near
the boundary. That leads to the main difference between this
situation and the one discussed in the previous section. The
equation for the average scalar turns to
 \begin{equation} \label{de2}
 s_0 q \partial_z \langle \theta \rangle
 = \left[\mu \partial_q q^4 \partial_q +
 \kappa \partial_q^2\right]\langle \theta \rangle.
 \end{equation}
Despite the additional factor $q$ in the left-hand side of the
equation in comparison with Eq. (\ref{pe11}), the qualitative
picture of scalar evolution remains the same. At the first stage
the scalar is mostly situated in the layer of the width $\delta =
s_0/(\mu z)$, and the molecular diffusion can be neglected. The
scalar decay at this stage is algebraic with the longitudinal
coordinate $z$. When $\delta$ reaches the boundary layer width
$r_{bl}$, the molecular diffusion becomes relevant, and the scalar
decay starts to be exponential.

As in the previous case, it is possible to obtain complete
statistical properties of the scalar at the first stage. In this
case one can omit the molecular diffusion term. Hence, averages of
any single-point functions $\Phi(\theta)$, such as the moments
$\langle \theta^n\rangle$ or the scalar PDF $P(\theta,q,z) =
\langle\delta[\theta-\theta(q,z)]\rangle$ are described by the
same equation
 \begin{equation}
 \partial_z \langle\Phi\rangle = \frac{\mu}{s_0 q}
 \partial_q q^4 \partial_q  \langle\Phi\rangle
 \equiv \hat{H} \langle\Phi\rangle.
 \label{pipea} \end{equation}
This is a linear equation, and it can be solved using the Green
function formalism. Namely, the solution of Eq. (\ref{pipea}) can
be written as
 \begin{equation}
 \langle \Phi(z,q)\rangle=
 \int_0^\infty\mathrm dq'\,G(z,q,q') \Phi(z=0,q') \,,
 \label{conv} \end{equation}
where $G$ is the Green function. In order to obtain an explicit
equation for the Green function one should first solve the
corresponding eigenvalue problems. Since the operator $\hat{H}$ is
not Hermitian, then one should find both it's right and left eigen
functions. They are solutions of the equations $\hat{H}f_\lambda +
\lambda f_\lambda =0$ and $\hat{H}^+ g_\lambda + \lambda g_\lambda
= 0$, respectively, where $H^+= (s_0/\mu)\partial_q q^4 \partial_q
q^{-1}$. We find
 \begin{eqnarray}
 f_\lambda = x^3 J_3(x),\quad
 g_\lambda = x J_3(x) , \quad
 x = 2\sqrt{{\lambda}/{q}}
 \end{eqnarray}
Using the orthogonality relation
 \begin{equation*}
 \int_0^\infty \mathrm{d} q\, g_\lambda(q)f_\nu(q)
 = 16 \lambda^2 \delta(\lambda-\nu)\,,
 \end{equation*}
one derives
 \begin{eqnarray}
 G(z,q,q') = \int_0^\infty
 \frac{\mathrm{d}\lambda}{16\lambda^2}
 \exp(-\lambda \delta^{-1})f_\lambda(q)g_\lambda(q')
 \nonumber \\
  = \frac{\delta}{q^{3/2}(q')^{1/2}}
 \exp\left(-\delta/q-\delta/q'\right)
 I_3\left(2\delta/\sqrt{qq'}\right) \,.
 \label{pipegreen} \end{eqnarray}

Since we assumed, that $\theta=\theta_0(q)$ initially, then for
the $n$-th order scalar moment $\langle\theta^n(q,z)\rangle$ the
initial condition is $\theta_0^n(q)$. If $\delta\ll L$ then one
can substitute $\langle\theta^n(0,z)\rangle$ by $\vartheta_0^n$,
where $\vartheta_0 = \theta_0 (0,0)$. In this case an explicit
integration in Eq. (\ref{conv}) leads to the expression
 \begin{equation}
 \langle\theta^n(z,q)\rangle=
 \frac{\vartheta_0^n\delta^3}{6q^3}\exp\left(-\delta/q\right)
 ~_1\!F_1\left(1,4,\delta/q\right) \,.
 \label{univa} \end{equation}
Again, we obtain a universal profile. If $q\gg\delta$ then both
the exponent and $~_1\!F_1$ can be substituted by unity to obtain
$\langle\theta^n(z,q)\rangle={\vartheta_0^n\delta^3}/(6q^3)$. If
$q\ll\delta$ then $\langle\theta^n(z,q)\rangle={\vartheta_0^n}$.
In order to obtain the scalar PDF, one should use the initial
distribution $P(z=0,q,\theta) = \delta(\theta -\theta_0(q)) =
-(1/\theta_0'(q))\delta(q - q_0(\theta))$. Here $q_0(\theta)$ is defined,
as previously, by the relation $\theta_0(q_0)=\theta$. The convolution
with the Green function is now
 \begin{eqnarray} && \label{pipepdf}
 P(z,q,\theta) = \int \mathrm{d}q' G(z,q,q') P(0,q,\theta)
 \\ && \nonumber
 = -\frac{1}{q^{3/2} q_0^{1/2}}\frac{\delta}{\theta_0'(q_0)}
 \exp\left(-\frac{\delta}{q}-\frac{\delta}{q_0}\right)
 I_3\left(2\frac{\delta}{\sqrt{q q_0}}\right) \,.
 \end{eqnarray}
At large distances, $q\gg \delta^2/L$ (and also $L\gg\delta$) one
can use the approximation:
 \begin{equation} \label{pipepdfapprox}
 P(z,q,\theta) \approx -\frac{\delta^4}{6 q^3 q_0^2
 \theta_0'(q_0)}\exp\left(-\delta/q\right)\,.
 \end{equation}
Again this expression cannot be exploited for calculation of
moments of the passive scalar because of the divergency near the
maximum value of $\theta$, $\vartheta_0$.

For the pipe, the characteristic length where the diffusion
becomes relevant is determined by the same expression (\ref{rbl}).
When $\delta$ diminishes down to $r_{bl}$ another regime comes.
For the density of pollutants the regime is characterized by an
exponential decay of the passive scalar moments along the pipe.
For the first moment the decrement of the decay can be extracted
from Eq. (\ref{de2}), it is determined by the smallest eigen value
of the operator in the right-hand side of the equation. Thus, for
large enough $z$ the decay law $\langle\theta\rangle\propto
\exp(-\alpha z)$ is correct where $\alpha\sim\kappa^{1/4}
\mu^{3/4}s_0^{-1}$. The eigenvalue problem, corresponding to the
equation (\ref{de2}) can be solved numerically, then one obtains
$\alpha\approx 3.72 \kappa^{1/4}\mu^{3/4}s_0^{-1}$.

For the temperature, fixed at the boundary, we obtain a
quasi-stationary distribution if the fixed value is
$z$-independent. The average temperature is described by the
equation (\ref{de2}) where $z$-derivative is dropped. Then we
return to the same equation as previously (for the decay in the
vessel), with the solution (\ref{rbl3}). Higher moments of the
temperature are slightly different than for the decay in the
vessel. However, qualitatively their behavior is the same. Namely,
$\langle\theta^n\rangle\approx \vartheta_0^n$ if $q\ll r_{bl}$,
and $\langle\theta^n\rangle\sim\vartheta_0^n (r_{bl}/q)^3$, if
$q\gg r_{bl}$.

One can examine the pair correlation function $F$ of the passive
scalar. Qualitatively, it is the same, as for the decay in the
vessel. Namely, the main anomalous dependence of the function is
related to the second moment of the passive scalar. The function
is practically undistinguished from the second moment if
$q_1,q_2\ll r_{bl}$. As to the dependence on the relative
separation between the points $\bm r_1$ and $\bm r_2$ for
$q_1,q_2\gg r_{bl}$, it is described in terms of a function
$\zeta$, $\left\langle\theta^2(Q)\right\rangle\,(1+\varsigma)$,
possessing a non-trivial scaling behavior with a self-similar
factor depending on the combination $Q\varrho/q$.

\section{Conclusion}

We have investigated the passive scalar (concentration of
pollutants or temperature) evolution in the chaotic and turbulent
flows near the boundary (wall), which dominates the advanced
stages of the passive scalar homogenization. There are some
universal features of the decay related to the universal velocity
dependence on the coordinate perpendicular to the wall. We assumed
that the Peclet number is large enough to guarantee the condition
$L\gg r_{bl}$ where $L$ is the width of the peripheral region and
$r_{bl}$ is the thickness of the boundary layer (which is
determined by equating the molecular diffusion and the turbulent
diffusion). The width of the peripheral region is the thickness of
the viscous boundary layer for the developed high-Reynolds
turbulence and is of the order of the vessel size for the elastic
turbulence. For the decay along a pipe $L$ is of the order of the
pipe radius.

We have considered the evolution beginning after that
homogenization of the passive scalar is finished in the bulk. Then
the initial passive scalar distribution is characterized by the
length $L$, which is much larger than $r_{bl}$. The first stage of
the passive scalar decay in the peripheral region is insensitive
to diffusion. The characteristic length of the passive scalar
distribution at this stage $\delta$ satisfies the inequalities
$L\gg \delta \gg r_{bl}$. For the decay in a closed vessel the
length behaves as $\delta\propto t^{-1/2}$, whereas for the decay
along pipe $\delta\propto z^{-1}$ (where $z$ is coordinate along
the pipe). The passive scalar inhomogeneity is kept mainly for
separations $q$ from the boundary, satisfying $q<\delta$ whereas
for larger $q$ the advection effectively homogenize the passive
scalar (the mechanism is in stretching producing small-scale
fluctuations of the passive scalar, which are effectively removed
by diffusion).

When $\delta$ achieves the thickness of the diffusion boundary
layer, $r_{bl}$, the decay of the concentration of pollutants
becomes exponential. This character of the decay is explained by
the space distribution of the passive scalar inhomogeneities,
which are concentrated mainly in the diffusive boundary layer, and
decay due to the flow to the bulk, which is proportional to the
level of the inhomogeneities. The decrements of the decay can be
expressed in terms of the Peclet number {which is
$\mathrm{Pe}=V_LL/\kappa$}. For the viscous boundary layer the
Peclet number coincides with the Schmidt number, whereas for the
elastic turbulence it differs from the Schmidt number by the
factor which is the Reynolds number. For the time decay the
decrement is proportional to $\mathrm{Pe}^{-1/2}$, whereas for the
decay along the pipe the decrement is proportional to
$\mathrm{Pe}^{-1/4}$. For the temperature fixed at the boundary
the second stage is characterized by a quasi-stationary
distribution, since the bulk serves as a big reservoir keeping its
temperature constant and absorbing heat, transferred from the
boundary.

Our theoretical predictions can be compared with the experimental
data \cite{01GS,03BGJSS}, obtained for the dye concentration decay
(homogenization) along the curvilinear pipe, where the chaotic
flow (elastic turbulence) is excited in a dilute polymer solution.
The most interesting comparison concerns the
$\mathrm{Pe}$-dependence of different quantities, which was
extracted from the experiment \cite{03BGJSS} by varying the
diffusion coefficient $\kappa$ by two orders of magnitude (it was
done by changing the molecular weight of the dye carriers). The
experimental data (corresponding to the second stage in our
theoretical scheme) are in a good agreement with the law
$\mathrm{Pe}^{-1/4}$ both for the thickness of the diffusion
boundary layer, $r_{bl}$, (which is determined experimentally by a
maximum in the passive scalar fluctuations) and for the decrements
of the passive scalar decay along the pipe.

The peripheral region serves as a source supplying the passive
scalar to the bulk. Since the passive scalar decay in the
peripheral region is slow (comparing to the bulk), it can be
considered as a quasi-stationary source of the passive scalar for
the bulk, where the passive scalar correlation functions adjust
adiabatically to the level of the supply. Therefore the situation
is close to one characteristic of the passive scalar, supplied by
pumping, in a smooth random velocity field. Statistical properties
of the passive scalar in this case were examined in the papers
\cite{59Bat,67Kra,95CFKLa}. Particularly, correlation functions of
the passive scalar in the bulk should possess a behavior close to
the logarithmic one. This prediction is also in agreement with the
experimental data \cite{01GS,03BGJSS}.

One of remarkable predictions of our theory is strong
intermittency of the passive scalar and extreme anomalous scaling
of its moments for the dependence on the separation from the
boundary $q$: all the moments scale identically as $q^{-3}$ for
$L\gg q\gg\delta$ or $L\gg q\gg r_{bl}$, depending on the stage.
Physically that means that the passive scalar decay is related to
rare events like escapes of separate passive scalar blobs from the
boundary layer with their subsequent transportation to the bulk.
As to the correlation functions of the passive scalar, our theory
predicts some definite self-similar and scaling behavior, which
are complex due to the space inhomogeneity characteristic of the
problem. It would be interesting to check the predictions
experimentally.

The final remark concerns an extension of our results to other
problems. As it was noted in the paper \cite{03CLb}, the scheme
developed for the passive scalar decay can be without serious
modifications applied to fast binary chemical reactions. We
believe that minor modifications of the scheme can make it
applicable for more complicated chemical reactions. The other
problem, which can be posed for the peripheral region, is dynamics
of polymers (say, for the case of the elastic turbulence). Then
one should go away the scope of the passive approach.

\acknowledgments

We thank M. Chertkov and I. Kolokolov for helpful discussions and
V. Steinberg for valuable remarks.

\onecolumngrid
\appendix

\section{pair correlation function}
\label{pairderiv}

The anomalous diffusion operator $\hat{A}$ in the dynamics of pair
correlation functions given by the last terms of the equation
(\ref{pe8}). It can be rewritten as
 \begin{equation}\label{an1}
 \hat{A} = D^{ij}\frac{\partial}{\partial X^i}\frac{\partial}{\partial X^j} +
 \left(\frac{\partial D^{ij}}{\partial X^i}\right)\frac{\partial}{\partial X^j}.
 \end{equation}
Here $X^i$ is extended set of coordinates $X =
\{q_1,r_{11},r_{12},q_2,r_{21},r_{22}\}$, and
\begin{equation}
 D = \left(
 \begin{array}{cccccc}
  \mu q_1^4 & 0 & 0 & H_3 q_1^2 q_2^2 & \mu q_1^2 q_2 (r_{11}-r_{21}) & \mu q_1^2 q_2(r_{12}-r_{22}) \\
  0 &  H_{10} q_1^2 & 0 & \mu q_1 q_2^2 (r_{21}-r_{11}) & H_{1} q_1 q_2 + d^{11} & d^{12}\\
  0 & 0 & H_{10} q_1^2 & \mu q_1 q_2^2 (r_{22}-r_{12}) & d^{21} & H_{1} q_1 q_2 +d^{22} \\
  H_3 q_1^2 q_2^2 & \mu q_1 q_2^2 (r_{22}-r_{12}) & \mu q_1 q_2^2 (r_{21}-r_{11}) & \mu q_2^4 & 0 & 0 \\
  \mu q_1^2 q_2 (r_{11}-r_{21}) & H_{1} q_1 q_2 + d^{11} & d^{12} & 0 & H_{10}q_2^2 & 0 \\
  \mu q_1^2 q_2(r_{12}-r_{22}) & d^{21} & H_{1} q_1 q_2 +d^{22} & 0 & 0 &  H_{10} q_2^2
 \end{array}
 \right)
\end{equation}
\begin{eqnarray}
 H_1 = H_{10} -(\mu+3 H_{20}/2)\left((r_{11}-r_{21})^2+(r_{12}-r_{22})^2\right)\\
 H_3 = \mu+H_{30}\left((r_{11}-r_{21})^2+(r_{12}-r_{22})^2\right) \\
 d^{ij} = H_{20} q_1 q_2 (r_{1i}-r_{2i})(r_{1j}-r_{2j})
\end{eqnarray}
Now we turn to the new variable set $Y = \{Q,q,R_1,R_2,\varrho,\phi\}$, such
that $Q = (q_1+q_2)/2,q=q_1-q_2,R_i=(r_{1i}+r_{2i})/2,\varrho\cos\phi =
r_{11}-r_{21},\varrho\sin\phi = r_{12}-r_{22}$. In this case, the anomalous
diffusion operator transforms to
\begin{equation}\label{an2}
 \hat{A} = D^{ij} T^k_i T^l_j\frac{\partial}{\partial Y^k}\frac{\partial}{\partial Y^l} +
 D^{ij} T^k_i\left(\frac{\partial T^l_j}{\partial
 Y^k}\right)\frac{\partial}{\partial Y^l}+
 \left(\frac{\partial D^{ij}}{\partial X^i}\right)T^k_j\frac{\partial}{\partial
 Y^k} \equiv \tilde{D}^{ij}\frac{\partial}{\partial
 Y^i}\frac{\partial}{\partial Y^j}+f^i\frac{\partial}{\partial Y^i},
\end{equation}
where $T^i_j={\partial Y^i}/{\partial X^j}$ is the Jacoby matrix:
 \begin{equation}
 T = \left(
 \begin{array}{cccccc}
  1/2 & 0 & 0 & 1/2 & 0 & 0 \\
  1 & 0 & 0 & -1 & 0 & 0 \\
  0 & 1/2 & 0 & 0 & 1/2 & 0 \\
  0 & 0 & 1/2 & 0 & 0 & 1/2 \\
  0 & \cos\phi & \sin \phi & 0 & -\cos\phi & -\sin\phi \\
  0 & -\sin(\phi)/\varrho & \cos(\phi)/\varrho & 0 & \sin(\phi)/\varrho &
  -\cos(\phi)/\varrho
 \end{array}
 \right)
 \end{equation}
After some algebraic manipulations one can easily obtain the new
diffusion tensor $\tilde{D}^{ij}$ and the first order term $f^i$.
The exact expressions are rather bulky. However, if the initial
distribution of scalar depends only on $q$, the only relevant
terms, are those, preceding
$\partial_Q,\partial_q,\partial_\varrho$. Therefore one can reduce
the final variables set to $\{Q,q,\varrho\}$. In the limit $Q\gg
q,\rho \to 0$  one has
\begin{equation}
 \tilde{D} = \left(\begin{array}{ccc}
  Q^4(\mu+H_{30}\varrho^2/2) &
  2\mu Q^3 q  & - \mu Q^3 \varrho  \\
  2\mu Q^3 q  & 4\mu q^2 Q^2- 2 H_{30} Q^4 \varrho^2
   & - \mu Q^2 q \varrho  \\
  - \mu Q^3 \varrho  & - \mu Q^2 q \varrho  &
  (2\mu+H_{20})Q^2+ H_{10} q^2
 \end{array}\right)
 \end{equation}
 \begin{eqnarray}
 f^Q = 4 Q^3(\mu + H_{30}\rho^2 /2), \qquad f^q = 0,\qquad f^\rho =
 \frac{H_{10} q^2}{\varrho} - 4 (\mu + 3 H_{20}/2) Q^2 \rho
 \end{eqnarray}
This yields the following expression for $\hat{A}$:
 \begin{eqnarray}
 \hat{A} = (\mu+H_{30}\rho^2/2) \partial_Q Q^4 \partial_Q + 4\mu Q^3 q
 \partial_Q\partial_q - 2\mu Q^3 \varrho \partial_Q\partial_\varrho +
 \nonumber \\
 +(4\mu Q^2 q^2 -2 H_{30} Q^4 \varrho^2)\partial_q^2 -2\mu Q^2 q \varrho
 \partial_q\partial_\varrho + \left[(2\mu + H_{20})Q^2\varrho^2+H_{10}
 q^2\right]\partial_\varrho^2+
 \nonumber \\
 +4 Q^3(\mu + H_{30}\rho^2 /2)\partial_Q+
 \left[\frac{H_{10} q^2}{\varrho} - 4 (\mu + 3 H_{20}/2) Q^2
 \rho\right]\partial_\varrho
 \end{eqnarray}

\twocolumngrid


\begin{thebibliography}{99}

\bibitem{00SS}
 B. I. Shraiman and E. D. Siggia,
 Nature (London) {\bf 405}, 639 (2000).

\bibitem{01FGV}
 G. Falkovich, K. Gaw\c{e}dzki, and M. Vergassola,
 Rev. Mod. Phys. {\bf 73}, 913 (2001).

\bibitem{03CLa}
 M. Chertkov and V. Lebedev,
 Phys. Rev. Lett. {\bf 90}, 034501 (2003).

\bibitem{03CLb}
 M. Chertkov and V. Lebedev,
 Phys. Rev. Lett. {\bf 90}, 134501 (2003).

\bibitem{MY}
 A. S. Monin and A. M. Yaglom, {\it Statistical Fluid Mechanics},
 MIT Press, Cambridge Mass. (1975).

\bibitem{00GS}
 A. Groisman and V. Steinberg, Nature {\bf 405}, 53 (2000).

\bibitem{59Bat}
 G. K. Batchelor, JFM {\bf 5}, 113 (1959).

 \bibitem{67Kra}
 R. H. Kraichnan, Phys. Fluids. {\bf 10}, 1417 (1967),
 JFM {\bf 47}, 525 (1971); Ibid {\bf 67}, 155 (1975).

\bibitem{95CFKLa}
 M. Chertkov, G. Falkovich, I. Kolokolov, and V. Lebedev,
 Phys. Rev. E {\bf 51}, 5609 (1995).

\bibitem{00EX}
 G. Eyink and J. Xin,
 J. Stat. Phys. {\bf 100}, 679 (2000).

\bibitem{Batchelor}
 G. K. Batchelor, {\it An Introduction to Fluid Dynamics},
 Cambridge University Press, 1967.

\bibitem{99Son}
 D. T. Son, Phys. Rev. E {\bf 59}, R3811 (1999).

\bibitem{99BF}
 E. Balkovsky and A. Fouxon, Phys. Rev. E {\bf 60}, 4164 (1999).

\bibitem{01GS}
 A. Groisman and V. Steinberg,
 Phys. Rev. Lett. {\bf 86}, 934 (2001); Nature {\bf 410}, 905 (2001).

\bibitem{Frisch}
 U.~Frisch, ``{\it Turbulence: the Legacy of A.~N.~Kolmogorov}",
 Cambridge University Press, New York (1995).

\bibitem{95KrM}
K.~Gawedzki and A.~Kupiainen, Phys.~Rev.~Lett. {\bf75}, 3608
(1995); B.~Shraiman and E.~Siggia, CRAS {\bf 321}, Ser. II, 279
(1995); M. Chertkov, G. Falkovich, I. Kolokolov and V. Lebedev,
Phys. Rev. E {\bf 52}, 4924 (1995).

\bibitem{73Lum}
 J.~L.~Lumley,
 Annu. Rev. Fluid Mech. {\bf 1}, 367 (1969);
  J. Polymer Sci.: Macromolecular Reviews
 {\bf 7}, 263 (1973).

\bibitem{BFL}
 E.~Balkovsky, A.~Fouxon, and V.~Lebedev,
 Phys. Rev. Lett. {\bf 84}, 4765 (2000);
 Phys. Rev. E {\bf 64}, 056301 (2001);
 A. Fouxon and V. Lebedev,
 Phys. Fluids {\bf 15}, 2060 (2003).

\bibitem{03BGJSS}
 T. Burghelea, E. Serge, and V.
 Steinberg, Mixing by polymers: experimental test of decay
 regime of mixing, Submitted to Phys. Rev. Lett., 2003.







\end{thebibliography}
\end{document}